\documentclass{elsart}

\usepackage{graphicx,hyperref}
\usepackage{epsfig,array}
\usepackage{booktabs}

\newcommand{\beq}{\begin{equation}}
\newcommand{\eeq}{\end{equation}}
\newcommand{\beqa}{\begin{eqnarray}}
\newcommand{\eeqa}{\end{eqnarray}}

\newcommand{\Tr}{\mathop{\rm Tr}}
\newcommand{\sla}[1]%
        {\kern .25em\raise.18ex\hbox{$/$}\kern-.75em #1}
\newcommand{\mybar}[1]%
        {\kern 0.8pt\overline{\kern -0.8pt#1\kern -0.8pt}\kern 0.8pt}

\newcommand\mycaption[1]{\caption{\footnotesize #1}} 

\begin{document} 

\begin{frontmatter}

\title{On the discretization of physical momenta in lattice QCD}

\author[romeII]{G.M.~de~Divitiis}
\author[romeII]{R.~Petronzio}
\author[romeII]{N.~Tantalo}

\address[romeII]{University of Rome ``Tor Vergata'' and INFN sez. RomaII, 
	Via della Ricerca Scientifica 1, 
	I-00133 Rome}

\begin{abstract}
The adoption of two distinct boundary conditions for two fermions species on a finite 
lattice allows to deal with arbitrary relative momentum  between the two particle 
species, in spite of the momentum quantization rule due to a limited physical box size.
We test the physical significance of this topological momentum by checking in the continuum limit 
the validity of the expected energy-momentum dispersion relations. 
\end{abstract}

\end{frontmatter}

\section{Introduction}
\label{sec:introduction} 

Among the restrictions of field theory formulations on a lattice, the finite volume
momentum quantization represents a severe limitation in various 
phenomenological applications. For example, in a two body hadron decay where the energies of the 
decay products, related by 4--momentum conservation to the masses 
of the particles involved,
cannot assume their physical values unless these masses are consistent 
with the momentum quantization rule. 
In this letter we propose a solution to the problem 
based on the use of different
boundary conditions for different 
fermion species\footnote{R.P. thanks M.~L\"uscher for drawing his attention on this point.}.

We test the idea in the simplest case of a flavoured quark-antiquark 
correlation used to determine asymptotically the energy of the corresponding meson.
In this case the fermion and the antifermion are the different fermion species and we show that
suitable different boundary conditions can propagate a meson with a momentum that 
can assume continuous values.

Section~\ref{sec:method} introduces the boundary conditions, 
section~\ref{sec:results} reports on the numerical 
results and section~\ref{sec:conclusions} summarizes the conclusions.

\section{Generalized boundary conditions}
\label{sec:method}

In order to explain the method to have continuous physical momenta
on a finite volume we first re--derive, for the sake of clarity, the momentum quantization rule in
the case of a particle with periodic boundary conditions (PBC).
To this end we consider a fermionic field $\psi(x)$ 
on a 4--dimensional finite volume of topology $T\times L^3$
with PBC in the spatial directions
\beq
\psi(x+\vec{e}_i\ L) = \psi(x)\;, \qquad i=1,2,3
\eeq
This condition can be re-expressed by Fourier transforming
both members of the previous equation
\beq
\int{d^4p\ e^{-ip(x+\vec{e}_i\ L)}\ \tilde{\psi}(p)}
= \int{d^4p\ e^{-ipx}\ \tilde{\psi}(p)}
\;, \qquad i=1,2,3
\label{eq:fspbc}
\eeq
It follows directly from the previous relation that, 
in the case of periodic boundary conditions, one has
\beq
e^{ip_iL} = 1 \quad \Longrightarrow \quad p_i = \frac{2\pi\ n_i}{L}
\;, \quad i=1,2,3  
\label{eq:pbcmom}
\eeq
where the $n_i$'s are integer numbers.
The authors of \cite{Jansen:1996ck} 
have first considered a generalized set of boundary conditions, that here
we call $\theta$--boundary conditions ($\theta$--BC),
depending upon the choice of a topological 3--vector $\vec{\theta}$
\beq
\psi(x+\vec{e}_i\ L) = e^{i\theta_i}\ \psi(x)\;, \qquad i=1,2,3
\label{eq:thetabc}
\eeq
The modification of the boundary conditions affects the zero of the 
momentum quantization rule. 
Indeed, by re-expressing equation~(\ref{eq:thetabc}) in Fourier space, 
as already done in the case of PBC in equation~(\ref{eq:fspbc}), one
has
\beq
e^{i(p_i-\frac{\theta_i}{L})L} = 1 \quad \Longrightarrow 
\quad p_i = \frac{\theta_i}{L} + \frac{2\pi\ n_i}{L}
\;, \quad i=1,2,3  
\label{eq:thetamom}
\eeq
It comes out that the spatial momenta are still quantized as for
PBC but shifted by an arbitrary \emph{continuous} 
amount ($\theta_i/L$).
The observation that this continuous
shift in the allowed momenta it is physical and can be thus profitably used in 
phenomenological
applications is the key point of the present work.  
The generalized $\theta$--dependent
boundary conditions of equation~(\ref{eq:thetabc}) can be  implemented by
making a unitary Abelian transformation on the fields satisfying $\theta$--BC
\beq
\psi(x) \quad \longrightarrow  \quad 
{\mathcal U}\ (\theta,x) \psi(x)= e^{-\frac{i \theta x}{L}}\  \psi(x)
\label{eq:unittransf}
\eeq
As a consequence of this transformation the resulting field satisfies
periodic boundary conditions but obeys a modified Dirac equation
\beqa
S[\bar{\psi},\psi] \quad
&\longrightarrow&  \quad \sum_{x,y}{ \bar{\psi}(x)\ {\mathcal U}(\theta,x)D(x,y) {\mathcal U}^{-1}(\theta,y)\ \psi(y)}
\nonumber \\
&=& \quad \sum_{x,y}{ \bar{\psi}(x) \ D_\theta(x,y)\ \psi(y)} 
\label{eq:Smod}
\eeqa    
where the $\theta$--dependent lattice Dirac operator $D_\theta(x,y)$ is obtained
by starting from the preferred discretization of the Dirac operator and 
by modifying the definition of the covariant lattice derivatives, i.e.
by passing from the standard forward and backward derivatives: 
\beqa
\nabla_\mu \psi(x) &=& \frac{1}{a}\left[U_\mu(x)\psi(x+a\ \hat{\mu}) - \psi(x) \right]
\nonumber \\
\nabla^\dagger_\mu \psi(x) &=& \frac{1}{a}\left[\psi(x) - U^{-1}_\mu(x-a\ \hat{\mu})\psi(x-a\ \hat{\mu}) \right]
\label{eq:standardder}
\eeqa    
to the $\theta$--dependent ones
\beqa
\nabla_\mu(\theta) \psi(x) &=& \frac{1}{a}\left[\lambda_\mu\ U_\mu(x)\psi(x+a\ \hat{\mu}) - \psi(x) \right]
\nonumber \\
\nabla_\mu(\theta)^\dagger \psi(x) &=& \frac{1}{a}\left[\psi(x) - \lambda_\mu^{-1}\ 
U^{-1}_\mu(x-a\ \hat{\mu})\psi(x-a\ \hat{\mu}) \right]
\label{eq:dermod}
\eeqa    
where we have introduced
\beq
\lambda_\mu = e^\frac{ia\theta_\mu}{L} \qquad \theta_0 = 0
\eeq

The authors of ref.~\cite{Bucarelli:1998mu} have considered for the first time
$\theta$--BC in perturbative phenomenological applications. They used the shift
in the momentum quantization rule, that they called a ``finite size momentum'',
in order to build an external source to probe the tensor structure 
of the Wilson operators. 
A similar analysis was then repeated non--perturbatively by the same group in ref.~\cite{Guagnelli:2003hw}.
The use of $\theta$--BC has been considered in different contexts 
also in~\cite{Bedaque:2004kc,Gross:1982at,Kiskis:2002gr,Kiskis:2003rd,Roberge:1986mm}.

In this work we point out that the term $\vec{\theta}/L$ acts as a true
physical momentum.

As a test, we calculate  the energy of a meson
made up by two different quarks with different  $\theta$--BC for the two 
flavours.
We work in the $O(a)$--improved Wilson--Dirac lattice formulation of the 
QCD within the Schr\"odinger Functional formalism~\cite{Luscher:1992an,Sint:1994un} but,
we want to stress that the use of $\theta$--BC in the spatial directions 
is completely decoupled from the choice
of time boundary conditions and 
can be profitably used outside the Schr\"odinger Functional formalism,
for example in the case of standard periodic time boundary conditions. 
Let us consider the following correlators
\beq
f_P^{ij}(\theta;x_0) = -\frac{a^6}{2}\sum_{\vec{y},\vec{z},\vec{x}}{
\langle\ \bar{\zeta}_i(\vec{y})\gamma_5 \zeta_j(\vec{z}) \ 
\bar{\psi}_j(x) \gamma_5 \psi_i(x) \ \rangle
} 
\label{eq:fp}
\eeq
where $i$ and $j$ are flavour indices, all the fields satisfy periodic boundary conditions 
and the two flavours obey different $\theta$--modified Dirac equations, as explained in equations~(\ref{eq:Smod}),
(\ref{eq:standardder})~and~(\ref{eq:dermod}). In practice it is adequate to choose the flavour $i$
with $\theta=0$, i.e. with ordinary PBC, and the flavour $j$ with $\theta\neq 0$.
After the Wick contractions the pseudoscalar correlator of
equation~(\ref{eq:fp}) reads
\beq
f_P^{ij}(\theta;x_0) = \frac{a^6}{2}\sum_{\vec{y},\vec{z},\vec{x}} \Tr{
\langle\ \gamma_5\ S_j(\theta;\vec{z},x)\ \gamma_5\ S_i(0;x,\vec{y})\ \rangle
} 
\label{eq:fpcontractions}
\eeq
where $S(\theta;x,y)$ and $S(0;x,y)$ are the inverse of the $\theta$--modified 
and of the standard Wilson--Dirac operators respectively.
Note that the projection on the momentum $\vec{\theta}/L$
of one of the quark legs in equation~(\ref{eq:fpcontractions}) it is not
realized by summing on the lattice points with an exponential factor but
it is encoded in the $\theta$--dependence of the modified Wilson--Dirac
operator and, consequently, of its inverse $S(\theta;x,y)$. 

This correlation is expected to decay exponentially at large times
as
\beq
f_P^{ij}(\theta;x_0) \qquad \stackrel{x_0\gg 1}{\longrightarrow} \qquad f_{ij}\  e^{- ax_0 E_{ij}(\theta,a)}
\eeq
where, a part from corrections proportional to the square of the lattice spacing, $E_{ij}$ is the
physical energy of the mesonic state
\beq
E_{ij}(\theta,a) = \sqrt{M_{ij}^2+\left(\frac{\vec{\theta}}{L}\right)^2} + O(a^2)
\label{eq:drlatt}
\eeq
here $M_{ij}$ is the mass of the pseudoscalar meson made of a $i$ and a $j$
quark anti--quark pair.
In the next section we will show the calculation of the meson energies for
different flavours and for different choices of $\vec{\theta}$. We will show that
after the continuum extrapolations we will find the expected relativistic
dispersion relations
\beq
E_{ij}^2 = M_{ij}^2+\left(\frac{\vec{\theta}}{L}\right)^2
\label{eq:drcont}
\eeq
%
%

\section{Numerical tests}
\label{sec:results}

All the results of this section are obtained in the quenched
approximation of the QCD.
We have done simulations on a physical volume 
of topology $T\times L^3$ with $T=2L$ and linear extension $L=3.2\ r_0$, where $r_0$
is a phenomenological distance parameter related
to the static quark anti--quark potential \cite{Sommer:1994ce}. In order
to extrapolate our numerical results to the continuum limit
we have simulated the same physical volume using three different
discretizations with number of points $(32\times 16^3)$, 
$(48\times 24^3)$ and $(64\times 32^3)$ respectively. 
We have fixed the three values of the bare couplings corresponding
to the different discretizations using the $r_0$ scale with the
numerical results given in \cite{Necco:2001xg}. All the parameters
of the simulations are given in table~\ref{tab:simpar}.
\begin{table}[t]
\begin{center}
\scriptsize
\begin{tabular}{cccccc}
\toprule               
$\bf \beta$ &          & $\bf L/a$   &          & $\bf k$      & $\bf r_0\ m^{RGI}$ \\
\toprule               
            & $\qquad$ &             & $\qquad$ &  0.132054    & 0.645(7)      \\
5.960       & $\qquad$ & 16          & $\qquad$ &  0.132609    & 0.520(6)      \\
            & $\qquad$ &             & $\qquad$ &  0.133315    & 0.362(5)      \\
            & $\qquad$ &             & $\qquad$ &  0.133725    & 0.269(4)      \\
\midrule               		  
            & $\qquad$ &             & $\qquad$ &  0.134208    & 0.655(9)      \\
6.211       & $\qquad$ & 24          & $\qquad$ &  0.134540    & 0.521(7)      \\
            & $\qquad$ &             & $\qquad$ &  0.134954    & 0.354(6)      \\
            & $\qquad$ &             & $\qquad$ &  0.135209    & 0.251(5)      \\
\midrule               		  
            & $\qquad$ &             & $\qquad$ &  0.134517    & 0.676(15)     \\
6.420       & $\qquad$ & 32          & $\qquad$ &  0.134764    & 0.540(12)     \\
            & $\qquad$ &             & $\qquad$ &  0.135082    & 0.365(10)     \\
            & $\qquad$ &             & $\qquad$ &  0.135269    & 0.262(9)      \\
\toprule
$\bf [\theta_x,\theta_y,\theta_z]$ & = & $[0.0,\ 0.0,\ 0.0]$ & $[1.0,\ 1.0,\ 1.0]$ & $[2.0,\ 2.0,\ 2.0]$ & $[3.0,\ 3.0,\ 3.0]$ \\
\toprule
\end{tabular}
\vskip 0.3cm
\mycaption{Parameters of the simulations. The values of the bare couplings has been chosen in order to
fix the extension of the physical volume $L = 3.2\ r_0$.
For each
value of the $k$ parameter we have simulated all the values of $\vec{\theta}$.}
\label{tab:simpar}
\end{center}
\vskip 1.5cm
\end{table}
The values of the RGI quark masses reported in table~\ref{tab:simpar} have been calculated
starting from the PCAC relation
\beq
m^{PCAC}_{ii} = \frac{ \tilde{\partial_0}f_A^{ii}(0;x_0) + a c_A \partial_0^\dagger \partial_0 f_P^{ii}(0;x_0)  }{2 f_P^{ii}(0;x_0)} 
\label{eq:PCACdiagonal}
\eeq
where $\partial_\mu$, $\partial^\dagger_\mu$ are the usual forward and backward lattice derivatives respectively while
$\tilde{\partial}_\mu$ is defined as  $(\partial_\mu + \partial_\mu^\dagger)/2$. The time correlator $f_P^{ij}(0;x_0)$ has 
already been defined in equation~(\ref{eq:fp}) while $f_A^{ij}(0;x_0)$ is defined in the following relation
\beq
f_A^{ij}(\theta;x_0) = -\frac{a^6}{2}\sum_{\vec{y},\vec{z},\vec{x}}{
\langle\ \bar{\zeta}_i(\vec{y})\gamma_5 \zeta_j(\vec{z}) \ 
\bar{\psi}_j(x) \gamma_0 \gamma_5 \psi_i(x) \ \rangle
} 
\label{eq:fa}
\eeq
The improvement coefficient $c_A$ has been computed non--perturbatively in \cite{Luscher:1997ug}.
The RGI quark masses are connected to the PCAC masses of equation~(\ref{eq:PCACdiagonal}) from
the following relation
\beq
m_{ii}^{RGI} = Z_M(g_0) \ \left[ 1 + (b_A-b_P)\ am_i \right] \ m^{PCAC}_{ii}(g_0)
\label{eq:rgimassii}
\eeq
where the renormalization factor $Z_M(g_0)$ has been computed non--perturbatively in \cite{Capitani:1998mq}.
Also the difference of the improvement coefficients $b_A$ and $b_P$ is known non--perturbatively from
\cite{deDivitiis:1998ka,Guagnelli:2000jw}. In (\ref{eq:rgimassii}) the masses $m_i$ are the bare ones
defined as
\beq
am_i = \frac{1}{2} \left[\frac{1}{k_i} - \frac{1}{k_c} \right] 
\label{eq:barequarkmass}
\eeq

For each value of the simulated quark masses reported in table~\ref{tab:simpar} we have
inverted the Wilson--Dirac operator for three non--zero values of $\vec{\theta}$.
Setting the lattice scale by using the physical value $r_0= 0.5$ fm, the expected values of the physical momenta
associated with the choices of $\vec{\theta}$ given in table~\ref{tab:simpar} are simply calculated according
to the following relation
\beq
|\vec{p}| \; = 
\; \frac{|\vec{\theta}|}{L} 
\; \simeq\; 0.125\ |\vec{\theta}| \; \mbox{GeV}
\quad=\quad \left\{
\begin{array}{l}
0.000 \\
0.217 \\
0.433 \\
0.650 \\
\end{array}  \right. \; \mbox{GeV} 
\qquad L \ \simeq\ 1.6\; \mbox{fm} 
\eeq
These values have to be compared with the value of the lowest physical momentum allowed
on this finite volume in the case of periodic boundary conditions, 
i.e. $|\vec{p}| \simeq 0.785$ GeV.
\begin{figure}[t]
\begin{center}
\epsfig{file=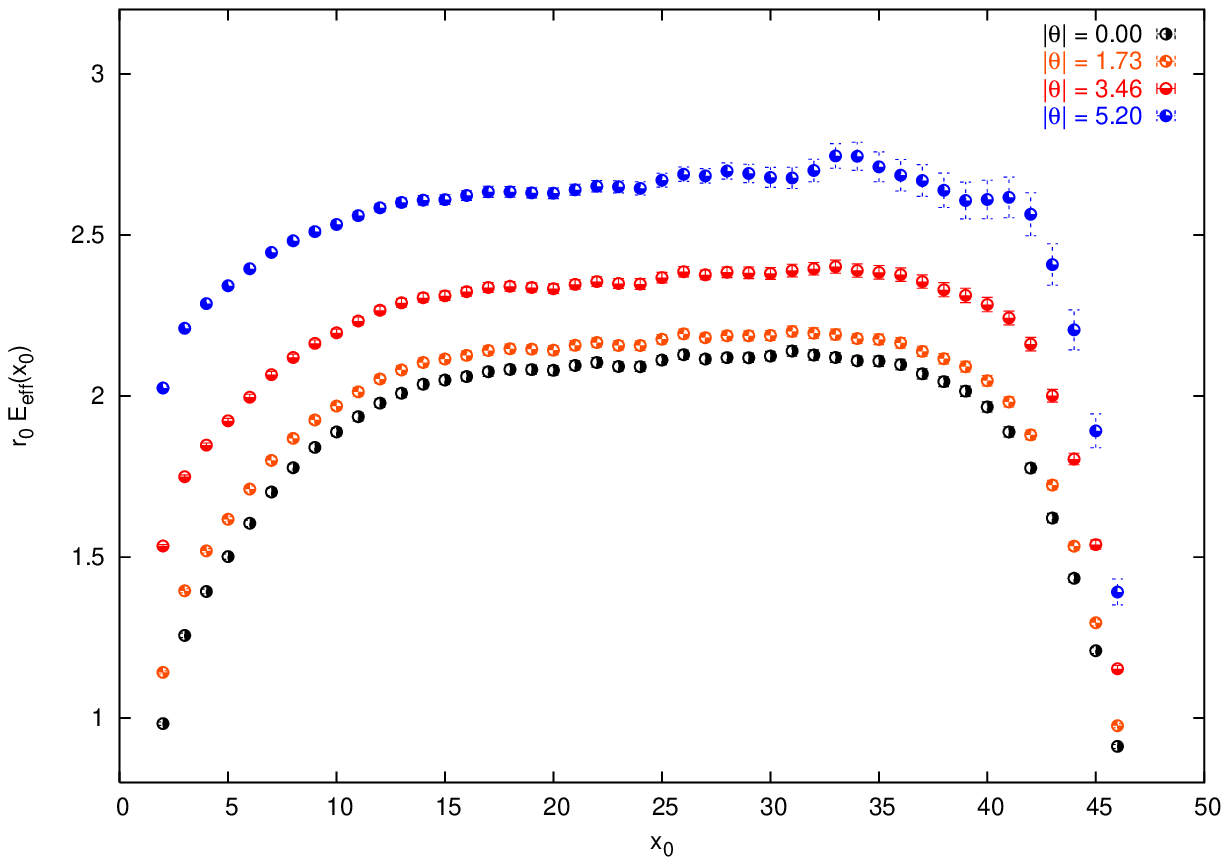,width=12cm}
\mycaption{Effective energies $E_{eff}^{ij}(\theta,a;x_0)$, as defined in eq.~(\ref{eq:effenerg})
at fixed cut--off. The results correspond to the simulation done at $\beta=6.211$
with $r_0\ m_1^{RGI} = 0.655$ and $r_0\ m_2^{RGI} = 0.354$.
Similar figures could have been shown for other combinations of the simulated quark masses
and for the other values of the bare coupling.}
\label{fig:effmass}
\end{center}
\vskip 1.5cm
\end{figure}

At fixed cut--off, for each combination of flavour indices and for each value of $\vec{\theta}$ 
reported in table~\ref{tab:simpar} we have extracted the 
effective energy from the correlations of eq.~(\ref{eq:fp}), $f_P^{ij}(\theta;x_0)$, as follows
\beq
a\ E_{eff}^{ij}(\theta,a;x_0) \; =\; \frac{1}{2}\ \log\left( \frac{f_P^{ij}(\theta;x_0-1)}{f_P^{ij}(\theta;x_0+1)} \right)
\label{eq:effenerg}
\eeq
In fig.~\ref{fig:effmass} we show this quantity for the simulation performed at $\beta = 6.211$
corresponding to $r_0\ m_1^{RGI} = 0.655$ and $r_0\ m_2^{RGI} = 0.354$, for each simulated value
of $\vec{\theta}$. As can be seen the correlations with higher values of $|\vec{\theta}|$ are
always greater than the corresponding ones with lower values of the physical momentum
\beq
|\vec{\theta}_1| > |\vec{\theta}_2| \quad \Rightarrow \quad  E_{eff}^{ij}(\theta_1,a;x_0) > E_{eff}^{ij}(\theta_2,a;x_0)
\label{eq:bounds}
\eeq
a feature that will be confirmed in the continuum limit.

In the continuum extrapolations we have fixed the physical values of the quark masses
slightly interpolating the simulated sets of numerical results.
\begin{figure}[t]
\begin{center}
\epsfig{file=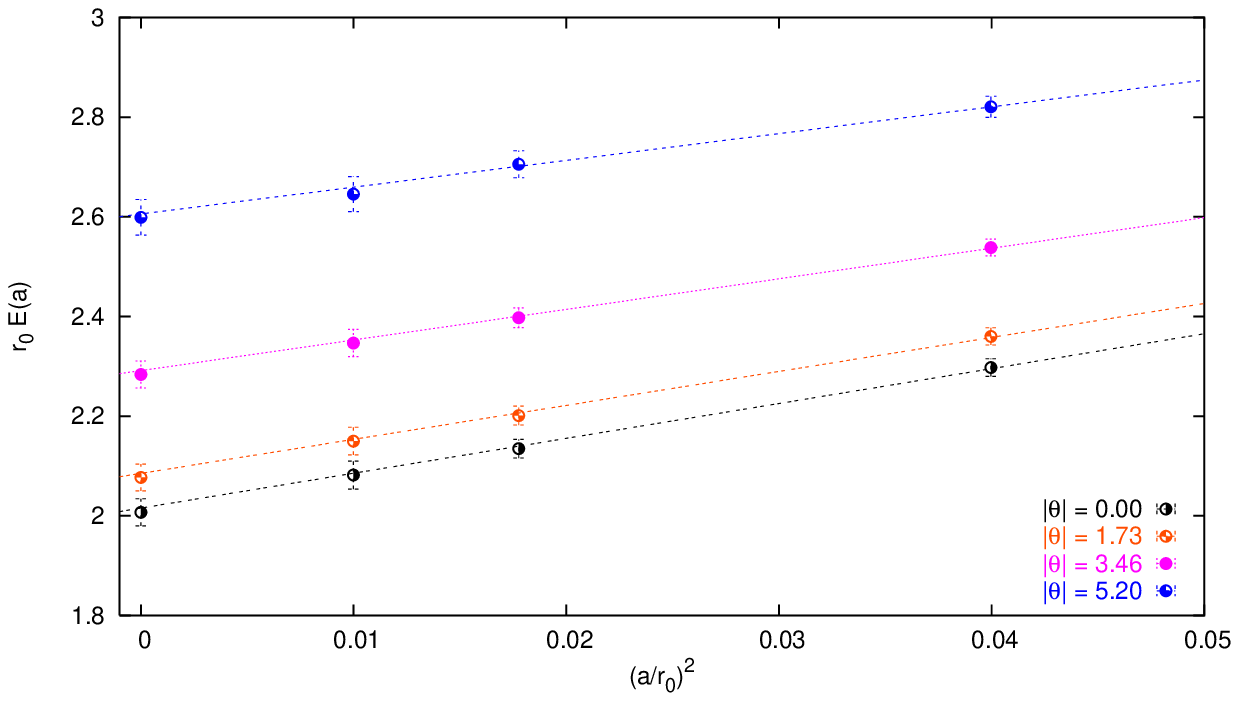,width=12cm}
\mycaption{Continuum extrapolations of the plateau averaged effective energies $E^{ij}(\theta,a)$. 
The results correspond to the quark masses $r_0\ m_1^{RGI} = 0.650$ and $r_0 m_2^{RGI} = 0.350$.
Similar figures could have been shown for other combinations of the simulated quark masses.}
\label{fig:contlimit}
\end{center}
\vskip 1.5cm
\end{figure}
Being interested in the ground
state contribution to the correlation of eq.~(\ref{eq:fp}), we have averaged the effective energies
over a ground state plateau of physical length depending upon the quark flavours.
We call $E^{ij}(\theta,a)$ the result of the average and in fig.~\ref{fig:contlimit}
we show a typical continuum extrapolation of this quantity. 
Similar figures could have been shown for the other values of simulated quark masses.
\begin{figure}[t]
\begin{center}
\epsfig{file=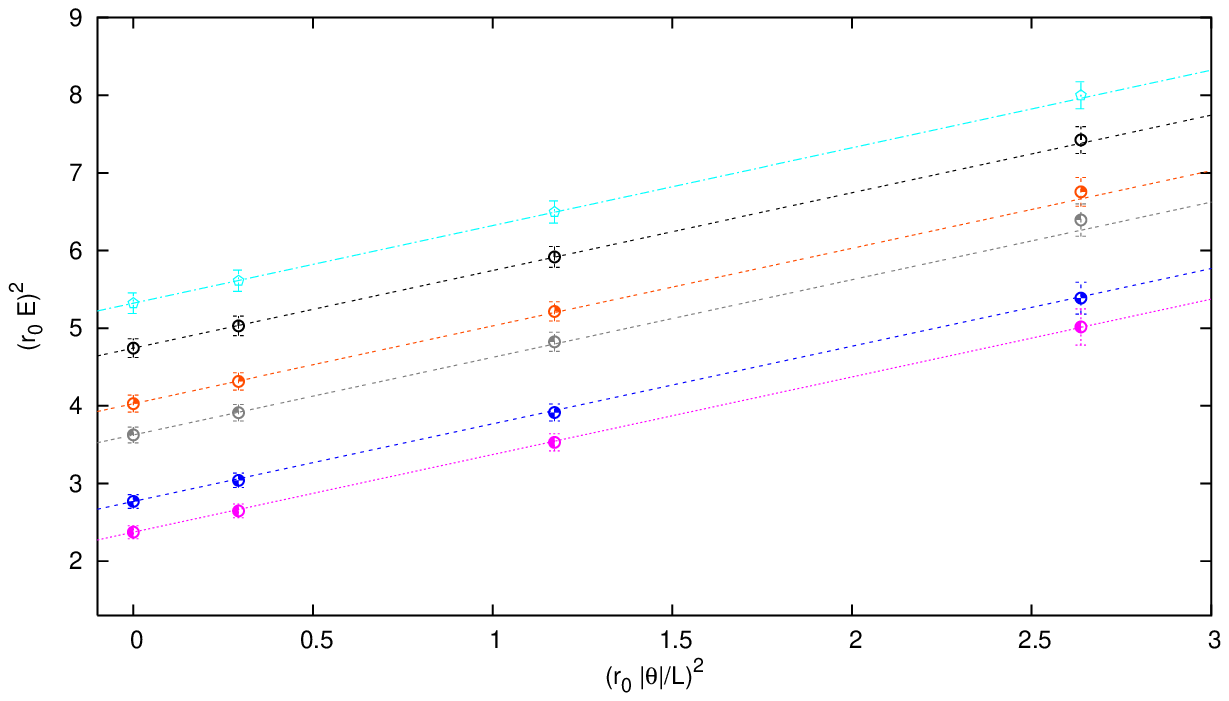,width=12cm}
\mycaption{
Continuum dispersion relations. The data correspond to different
combinations of the simulated quark masses and reproduce very well
the expected theoretical behavior, i.e. straight lines having
as intercepts the meson masses and as angular coefficients one (see eq.~\ref{eq:drcont}).
}
\label{fig:disprel}
\end{center}
\vskip 1.5cm
\end{figure}

The continuum results verify very well the dispersion relations of
equation~(\ref{eq:drcont}) as can be clearly seen from fig.~\ref{fig:disprel} in which
the square of $E^{ij}(\theta)$ for various combinations of the flavour indices is plotted
versus the square of the physical momenta $|\vec{\theta}|/L$. 
The plotted lines have not been fitted but have been obtained by using as
intercepts the simulated meson masses and by fixing their 
angular coefficients to one.

\clearpage

\section{Conclusions}
\label{sec:conclusions}

We have argued that the limitation
represented by the finite volume momentum quantization rule can be overcame
by using different boundary conditions for different fermion species.   

We have supported this observation by calculating the relativistic dispersion
relations satisfied by a set of pseudoscalr mesons in the case of quenched
lattice QCD. 
We have shown that the physical momentum carried by these particles
can be varied \emph{continuously} by enforcing different $\theta$--boundary conditions
(see eq.~(\ref{eq:thetabc}))  for the two quarks inside the mesons.

The method proposed can be applied to study all the
quantities of phenomenological interest that would benefit from the
introduction of continuous physical momenta like, for example, weak matrix elements. 
The suggestion can be applied in quenched QCD also in the case of flavourless mesons 
while can be extended to full QCD in the flavoured case only.

\begin{ack}
We warmly thank M.~L\"uscher for enlightening discussions.
We also thank F.~Palombi for useful remarks.
\end{ack}

\bibliographystyle{h-elsevier} 
\bibliography{tt}

\end{document}